\documentstyle[aps,prl,epsfig,floats]{revtex}

\begin{document}
\draft
\wideabs{

\title{Holon edge states in organic conductors 
(TMTTF)$\bf_2$X in the charge-gap regime}

\author{Andrei V. Lopatin and Victor M. Yakovenko}

\address{Department of Physics and Center for Superconductivity
   Research, University of Maryland, College Park, MD 20742-4111}

\date{\bf cond-mat/0106516, June 25, 2001}

\maketitle

\begin{abstract}
Using the bosonization technique, which separates charge and spin
degrees of freedom, we study a possibility of formation of the holon
edge states in a one-dimensional electron system with an energy gap in
the charge sector.  The results are applied to the
quasi-one-dimensional organic conductors (TMTTF)$_2$X.  The different
roles of the bond and site dimerizations in this material are
discussed.  We predict that the holon edge states should appear below
the temperature of the recently discovered ferroelectric transition,
where the inorganic anions X displace to asymmetric positions.
\end{abstract}
}

When the bulk electron states have an energy gap, additional states
may exist on the surface with the energies inside the gap.  Such
surface states were studied for a long time in the cases where the gap
has a band-structure origin \cite{Davison}.  More recently, edge
states attracted much interest in the cases where the gap originates
from a phase transition in the electron system.  Examples are midgap
states in the high-$T_c$ $d$-wave cuprate superconductors \cite{Hu}
and chiral states in the $p$-wave superconductor $\rm Sr_2RuO_4$ (see,
for example, Ref.\ \cite{Yakovenko-SrRu} and references therein).
Edge states in the quasi-one-dimensional (Q1D) organic conductors
(TMTSF)$\rm _2$X \cite{TMTSF} were studied for the triplet
superconducting state \cite{Sengupta-SC} and for the
magnetic-field-induced spin-density-wave (FISDW) state
\cite{Sengupta-FISDW}, which exhibits the quantum Hall effect.  In all
of these case, noninteracting electrons were subject to a mean-field
potential, so the problem reduced to a single-particle
Schr\"odinger-like equation with a boundary.  However, it is known
that a one-dimensional (1D) system of interacting electrons can be
described in terms of separate charge and spin sectors (see, for
example, review \cite{Firsov}).  An edge state in such a system with a
gap in the spin channel was studied in Ref.\ \cite{Fabrizio}.  In this
Letter, we use the technique developed in Ref.  \cite{Fabrizio} (see
also Ref.\ \cite{Gogolin}) to investigate another physical situation
where a gap opens in the charge sector.  The charge gap appears when
the electron system is commensurate with the crystal lattice, and the
umklapp interaction becomes relevant.  This is the case for the
(TMTTF)$_2$X family of Q1D organic conductors \cite{Yamaji}, which are
the best known examples of spin-charge separation. In these materials,
electric transport is thermally-activated due to the charge gap,
whereas spin susceptibility only weakly depends on temperature, which
indicates gapless spin sector.

\begin{figure}[b]
\centerline{\psfig{file=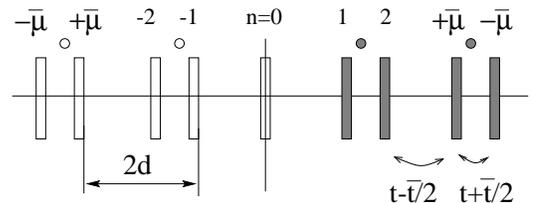,width=0.8\linewidth,angle=0}}
\caption{ A sketch of the (TMTTF)$_2$X chain. The vertical rectangles
  represent the organic molecules TMTTF, and the circles indicate the
  inorganic anions X.  The filled objects ($n=0,1,\ldots$) show the
  actual chain, whereas the open objects ($n=0,-1,-2,\ldots$)
  represent an artificial expansion of the system by the mirror
  reflection around the site $n=0$.}
\label{fig:chain}
\end{figure}

The (TMTTF)$_2$X crystals consist of parallel 1D molecular chains
quarter-filled with holes.  The umklapp term appears due to
dimerization of the chains.  Generally, there are two sources of
dimerization, one due to nonequivalence of hopping integrals along the
chain, and another due to nonequivalence of the site potentials.  The
unit cell consists of two organic molecules TMTTF and one inorganic
anion X.  The anions X are located outside of the TMTTF chain, between
the pairs of TMTTF molecules, as shown in Fig.\ \ref{fig:chain}.
Because of the anions, the bonds between the TMTTF molecules alternate
between short and long.  At high temperatures, the anions occupy
symmetric positions, so the TMTTF molecules themselves are equivalent.
However, it was recently proved that below a certain transition
temperature, the anions displace to asymmetric positions, making the
two TMTTF molecules in a unit cell nonequivalent \cite{Monceau}.  In
this Letter, we show that one source of dimerization controls the
existence and the localization length the edge states, whereas the
other source determines their energy.

In our theoretical model, we neglect interchain coupling and consider
just one semi-infinite chain.  For simplicity, we will refer to holes
as electrons and set $\hbar=1$.  The electron band structure of the
chain is described by the Hamiltonian
\begin{eqnarray} 
   & \hat H=-\sum\limits_{n=1,\,s}^{\infty}\left[t_{n}^{}
   (\hat\psi^\dagger_{n,s} \hat\psi_{n+1,s}^{} + {\rm H.c.})
   + \mu_{n}^{} \hat\psi^\dagger_{n,s} 
   \hat\psi_{n,s}^{} \right], &
\label{hamilton} \\
   & t_{n}=t-(-1)^n\bar t/2, \qquad \mu_{n}=\mu-(-1)^n\bar\mu, &
\label{t-mu}
\end{eqnarray}
where the index $n$ labels the TMTTF sites, $s=\uparrow,\downarrow$ is
the spin index, and $\hat\psi^{(\dagger)}_{n,s}$ are the creation and
annihilation operators of electrons.  The mean values of the intersite
tunneling amplitude and the chemical potential are $t$ and $\mu$,
whereas $\bar t$ and $\bar\mu$ are their staggering parts produced by
dimerization of the chain bonds and sites, respectively.

\begin{figure}
\centerline{\psfig{file=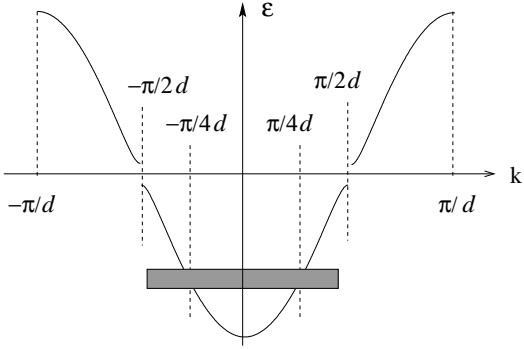,width=0.8\linewidth,angle=0}}
\caption{The energy dispersion for the Hamiltonian (\ref{hamilton}).
  The single-particle gap at $k=\pm\pi/2d$ opens due to backscattering
  (\ref{bs}) caused by the chain dimerization (\ref{t-mu}).  The
  correlation gap at $k=\pm\pi/4d$ is caused by umklapp interaction.}
\label{fig:dispersion}
\end{figure}

The energy dispersion for the Hamiltonian (\ref{hamilton}) and
(\ref{t-mu}) is sketched in Fig.\ \ref{fig:dispersion}.  For a given
chemical potential $\mu$, there exist two Fermi points $\pm k_F$.  To
study low-energy properties of the system, we introduce the operators
$\hat R$ and $\hat L$ describing the right- and left-moving electrons
with the momenta close to $+k_F$ and $-k_F$:
\begin{equation}  
\label{slowfields}
   \hat\psi_{s}(x)=\left[e^{ik_Fx}\,\hat R_s(x)
   -e^{-ik_Fx}\,\hat L_s(x)\right]\sqrt{d}.
\end{equation}
Here $x=nd$ is the continuous coordinate, and $d$ is the average
distance between the TMTTF sites.

The considered system (\ref{hamilton}) extends in the positive
direction from the site $n=1$ to a length $l$, which will be taken to
infinity.  The wave functions $\psi_n$ can be expanded to include the
site $n=0$ with the vanishing boundary condition $\psi_0=0$.  Then the
system can be artificially expanded from the interval $(0,l)$ to
$(-l,l)$ by the mirror reflection around the origin $n=0$, as shown in
Fig.\ \ref{fig:chain}, while keeping only the antisymmetric wave
functions $\psi_n=-\psi_{-n}$.  Because the Hamiltonian
(\ref{hamilton}) couples only the nearest neighboring sites, such an
expanded problem is equivalent to the original semi-infinite problem,
because there is no direct coupling between the intervals $(-l,0)$ and
$(0,l)$ other than through the point $n=0$, where the wave function
vanishes.  As Fig.\ \ref{fig:chain} shows, the mirror reflection
changes the sign of the bond dimerization parameter $\bar t$, but
keeps the same sign for the site dimerization $\bar\mu$.  Thus, for
the expanded problem, Eq.\ (\ref{t-mu}) should be replaced by
$t_n=t-{\rm sgn}(n)(-1)^n\,\bar t$.  In the continuous expanded
system, the boundary condition $\hat\psi(0)=0$ is satisfied by
demanding that $\hat\psi(x)=-\hat\psi(-x)$.  Using Eq.\ 
(\ref{slowfields}), the latter condition gives
\begin{equation}
  \hat R(x)=\hat L(-x).
\label{relation}
\end{equation}

\paragraph*{Dimerized chain at half filling.}  
As a warm-up, let us study the edge states for noninteracting
electrons at half filling, where $\mu=0$ and $k_F=\pi/2d$ (see Fig.\
\ref{fig:dispersion}).  Although the model (\ref{hamilton}) and
(\ref{t-mu}) at half filling does not correspond to (TMTTF)$_2$X, it
does describe some 1D polymers \cite{Braz-polymer}, which have both
site and bond dimerization.  The nondimerized part of the Hamiltonian
(\ref{hamilton}) is
$$
   \hat H_0=\sum_s \int\limits_0^l dx\, v_F\left[-\hat
   R_s^\dagger(x)\,i\partial_x \hat R_s(x) + \hat
   L_s^\dagger(x)\,i\partial_x \hat L_s(x)\right],
$$
where $v_F=2td$ is the Fermi velocity.  The dimerized part of the
Hamiltonian (\ref{hamilton}) gives the backscattering term:
\begin{equation}
   \hat H_{\rm bs}=\sum_s \int_0^l dx \left[(-\bar\mu + i\bar t)\,
   \hat R_s^\dagger(x)\hat L_s(x) + {\rm H.c.}\right].
\label{bs}
\end{equation}
The energies $\varepsilon$ and the wave functions $R(x)$ and $L(x)$ of
electron eigenstates can be found from the corresponding Schr\"odinger
equation
\begin{eqnarray}
   \left(\begin{array}{cc} 
   -iv_{\rm F}\partial_x & -\bar\mu + i\bar t \\ 
   -\bar\mu - i\bar t & +iv_{\rm F}\partial_x 
   \end{array}\right) {R(x) \choose L(x)} 
   = \varepsilon {R(x) \choose L(x)}.
\label{matrix}
\end{eqnarray}
Eq.\ (\ref{matrix}) gives the spectrum
$\varepsilon=\sqrt{v_F^2k^2+\Delta^2}$ for the bulk states with the
energy gap $\Delta=\sqrt{\bar\mu^2+\bar t^2}$.

With the boundary condition $\psi(0)=0$, which translates into
$R(0)=L(0)$ using Eq.\ (\ref{slowfields}), Eq.\ (\ref{matrix}) also
gives a solution localized at the edge (the Shockley state
\cite{Davison}):
\begin{equation}
   R_e(x)=L_e(x)=e^{-\kappa x},\quad \varepsilon_e=-\bar\mu,
\quad\kappa=-\bar t/v_F. 
\label{E_e}
\end{equation}
Eq.\ (\ref{E_e}) shows that the energy of the edge state is determined
by the site dimerization $\bar\mu$, whereas its localization length
$1/\kappa$ by the bond dimerization $\bar t$.  The edge state exists
only when $\bar t<0$, i.e.\ when the edge site ($n=1$) is connected to
the next site ($n=2$) by the long bond.  That configuration is
opposite to what is drawn in Fig.\ \ref{fig:chain}, where the first
bond is short.

Using Eq.\ (\ref{relation}), we get Eq.\ (\ref{matrix}) in the
expanded picture:
\begin{equation}  
\label{onlyright}
   -i\partial_xR(x)-[\bar\mu-i\bar t\,{\rm sgn}(x)]\,R(-x)
   =\varepsilon R(x).
\end{equation}
The edge state (\ref{E_e}) maps onto a state localized around the kink
in the imaginary part of the potential in Eq.\ (\ref{onlyright}),
which is produced by the sign change of the bond dimerization $\bar t$
upon mirror reflection around $n=0$.  Such soliton states were studied
for polymers in Ref.\ \cite{Braz-polymer}.

\paragraph*{Quarter filling.} 
In this case, $k_F=\pi/4d$, as shown in Fig.\ \ref{fig:dispersion},
and there is no single-particle energy gap at the Fermi level.
However, a correlation gap (shown as the shaded bar in Fig.\
\ref{fig:dispersion}) may develop at the Fermi level due to
interaction between electrons.  Thus, we add the Hubbard interaction
$U\sum_n \hat\psi^\dagger_{n\uparrow} \hat\psi^\dagger_{n\downarrow}
\hat\psi_{n\downarrow}^{} \hat\psi_{n\uparrow}^{}$.  It can be
rewritten in the standard $g$-ology form \cite{Firsov} in terms of the
slow fields $\hat R$ and $\hat L$.  Without dimerization, one obtains
\begin{eqnarray}
   \hat H_{\rm int}=\sum_{s,s^\prime} \int_0^l dx \left\{
   2( g_{1\|}\delta_{ss'}+g_{1\perp}\delta_{\bar ss'}) 
   \hat R_s^\dagger \hat L_{s'}^\dagger \hat R_{s'}^{} \hat L_s^{}
   \right.
\nonumber \\ 
   \left. + 2g_2\,
   \hat R_s^\dagger \hat L_{s'}^\dagger \hat L_{s'}^{} \hat R_s^{} 
   + g_4\,[
   \hat R_s^\dagger \hat R_{s'}^\dagger \hat R_{s'}^{} \hat R_s^{}
   + (\hat R\rightarrow\hat L) ] \right\},
\label{g-ology}
\end{eqnarray}
where $\bar s=\downarrow,\uparrow$ for $s=\uparrow,\downarrow$.
 
Introducing the densities of spin and charge
\begin{equation}  
\label{sc}
   \hat\rho_R=\frac{\hat R_\uparrow^\dagger \hat R_\uparrow^{}
   + \hat R_\downarrow^\dagger \hat R_\downarrow^{}}{\sqrt{2}},\quad
   \hat \sigma_R=\frac{\hat R_\uparrow^\dagger \hat R_\uparrow^{}
   - \hat R_\downarrow^\dagger \hat R_\downarrow^{}}{\sqrt{2}},
\end{equation}
the interaction Hamiltonian (\ref{g-ology}) can be written as
\begin{eqnarray} 
   \hat H_{\rm int} &=& \hat H_{int}^{(\sigma)} + \hat H_{int}^{(\rho)} 
   +2g_{1\perp}\sum_s\int_0^l dx \,
   \hat R_s^\dagger \hat L_{\bar s}^\dagger \hat R_{\bar s}^{}  \hat L_s^{},
\label{Hint} \\
   \hat H_{\rm int}^{(\rho)} &=& \int_0^l dx \left[
   g_4(\hat\rho_R^2+\hat\rho_L^2)-2g_5\,\hat\rho_R \hat\rho_L\right],
\label{H-rho} \\
   \hat H_{\rm int}^{(\sigma)} &=& -\int_0^l dx 
   \left[g_4(\hat\sigma_R^2 + \hat\sigma_L^2) +
   2g_{1\|}\,\hat\sigma_R \hat\sigma_L\right],
\label{H-sigma}
\end{eqnarray}
where $g_5=g_{1\|}-2g_2$.

\paragraph*{The umklapp term.} 
Taking into account that dimerization (\ref{t-mu}), treated as a
perturbation, hybridizes the states with the momenta $k=\pm
k_F=\pm\pi/4d$ and $k=\mp 3\pi/4d$, we make the following replacement
in Eq.\ (\ref{slowfields})
\begin{equation}  
\label{psi+pf}
   e^{\pm ik_Fx} \quad \longrightarrow \quad e^{\pm i\pi n/4}-
   \frac{\bar\mu \pm i\bar t/\sqrt{2}}{\delta E}
   \: e^{\mp i 3\pi n/4},
\end{equation}
where $\delta E=2t\sqrt{2}$.  Inserting Eqs.\ (\ref{slowfields}) and
(\ref{psi+pf}) into the Hubbard interaction term, we obtain the
umklapp term
\begin{eqnarray}  
   & \hat H_{\rm um}=\int_0^l dx
   \left[(g_3^a+ig_3^b)\, \hat R_\uparrow^\dagger \hat R_\downarrow^\dagger
   \hat L_\downarrow^{} \hat L_\uparrow^{}  + {\rm H.c.} \right], &
\label{Hum} \\
   & g_3^a= -\sqrt{2}\,U\bar\mu d/t, \quad g_3^b=U\bar td/t. &
\label{g_3}
\end{eqnarray}
The real part $g_3^a$ of the umklapp amplitude is proportional to the
site dimerization energy $\bar\mu$, which occurs due to the anion
displacement, and the imaginary part $g_3^b$ is proportional to the
bond dimerization $\bar t$.

\paragraph*{Bosonization with a boundary.}
As Eq.\ (\ref{relation}) shows, the problem on the expanded interval
$(-l,l)$ can be formulated in terms of the right-moving electrons
only.  The kinetic energy $\hat H_0$ becomes
\begin{equation}  
\label{freemodel}
   \hat H_0=-v_F\sum_s\int_{-l}^l dx\, 
   \hat R_s^\dagger(x)\,i\partial_x \hat R_s^{}(x),
\end{equation}
where $v_F=td/\sqrt{2}$ for the quarter filling.  Imposing the
periodic boundary condition $\hat R(-l)=\hat R(l)$ together with Eq.\ 
(\ref{relation}) results in the vanishing boundary condition on
$\hat\psi$ at the other end of the chain: $\hat\psi(l)=0$.  Now we can
use the standard bosonization technique \cite{Firsov} for $\hat R$.
The fermion fields $\hat R_s(x)$ can be represented in terms of the
Bose fields $\hat\theta_s(x)$:
\begin{equation}  
\label{R}
   \hat R_s(x)=\hat\eta_s\,e^{i\hat\theta_s(x)}\,(2\pi\zeta)^{-1/2},
\end{equation}
where $\zeta$ is a short-range cut-off distance of the order of $d$,
and $\hat\eta_s$ are the Majorana fermions:
$\{\hat\eta_\uparrow,\hat\eta_\downarrow\}=0$, $\hat\eta_s^2=1$.  The
Bose fields obey the commutation relations
\begin{equation}
\label{commutator}
   [\hat\theta_s(x),\hat\theta_{s'}(x')]=i\pi\,\delta_{ss'}\,
   {\rm  sgn}(x-x').
\end{equation}
Using the linear combinations
\begin{equation}
\label{ud}
   \hat\theta_\rho=(\hat\theta_\uparrow+\hat\theta_\downarrow)
   /\sqrt{2}, \quad
   \hat\theta_\sigma=(\hat\theta_\uparrow-\hat\theta_\downarrow)
   /\sqrt{2},
\end{equation}
the charge and spin densities (\ref{sc}), can be expressed in terms of
the charge and spin Bose fields (\ref{ud}):
\begin{equation}
   \hat\rho_{R}(x)=\partial_x\hat\theta_\rho(x)/2\pi, \quad
   \hat\sigma_{R}(x)=\partial_x\hat\sigma_\rho(x)/2\pi.
\label{rs}
\end{equation}

The Hamiltonian (\ref{freemodel}) can be rewritten in terms of the
densities (\ref{rs}) \cite{Firsov}:
\begin{equation}
\label{H0}
   \hat H_0=\pi v_F\int_{-l}^l dx\,
   \left[\hat\rho_R^2(x)+\hat\sigma_R^2(x)\right].
\end{equation}
Substituting Eqs.\ (\ref{relation}) and (\ref{R}) into the interaction
Hamiltonian (\ref{Hint}) and (\ref{Hum}) and adding the kinetic energy
(\ref{H0}), we get the full Hamiltonian of the system in the bosonized
form.  It takes the form $\hat H_\rho + \hat H_\sigma$, where the spin
and charge sectors are separated.  The charge Hamiltonian is
\begin{eqnarray}
  \hat H_\rho &=& \int_{-l}^l dx \left[\pi v_\rho\,\hat\rho_R^2(x)
    -g_5\,\hat\rho_R(x)\hat\rho_R(-x) \right.
\nonumber  \\
  &-& \left. \frac{g_3^a+ig_3^b\,{\rm sgn}(x)}{(2\pi\zeta)^2}\,
    e^{-i\sqrt{2}\,[\hat\theta_\rho(x)-\hat\theta_\rho(-x)]} \right].
\label{charge1}
\end{eqnarray}
Notice that the imaginary part of the umklapp amplitude in Eq.\ 
(\ref{charge1}) has a kink, because it is proportional to $\bar t$
(\ref{g_3}), which changes sign between $x<0$ and $x>0$.  The spin
Hamiltonian is
\begin{eqnarray}
   \hat H_\sigma &=& \int_{-l}^l dx \left[\pi v_\sigma\,\hat\sigma_R^2(x)
   -g_{1\parallel} \hat\sigma_R(x) \hat\sigma_R(-x) \right.
\nonumber \\
\label{spin1}
   &+& \left. \frac{2g_{1\perp}}{(2\pi\zeta)^2}\,
   e^{-i\sqrt{2}\,[\hat\theta_\sigma(x)-\hat\theta_\sigma(-x)]} \right].
\end{eqnarray}
In Eqs.\ (\ref{charge1}) and (\ref{spin1}), $v_{\rho,\sigma}=v_F\pm
g_4/\pi$.

The quadratic parts of the Hamiltonians (\ref{charge1}) and
(\ref{spin1}) can be diagonalized by the Bogoliubov transformations
\begin{eqnarray}
   && \hat\theta_{\rho,\sigma}(x) =
   \tilde\theta_{\rho,\sigma}(x) \cosh\phi_{\rho,\sigma}
   - \tilde\theta_{\rho,\sigma}(-x) \sinh\phi_{\rho,\sigma},
\\
   && \hat\rho_R(x) = \tilde\rho(x) \cosh\phi_\rho
   + \tilde\rho(-x) \sinh\phi_\rho,
\\
   && \hat\sigma_R(x) = \tilde\sigma(x) \cosh\phi_\sigma
   + \tilde\sigma(-x) \sinh\phi_\sigma,
\end{eqnarray}
where $\tanh2\phi_\rho =g_5/\pi v_\rho$ and
$\tanh2\phi_\sigma=g_{1\|}/\pi v_\sigma$.  The charge Hamiltonian
becomes
\begin{eqnarray}
   \hat H_\rho &=& \int_{-l}^l dx \left[\pi\tilde v_\rho\,
   (\partial_x\tilde\theta_\rho(x)/2\pi)^2 \right.
\nonumber \\ 
\label{charge}
   &-& \left. \frac{g_3^a+ig_3^b\,{\rm sgn}(x)}{(2\pi\zeta)^2}\,
   e^{-i\sqrt{2K_\rho}\,[\tilde\theta_\rho(x)-\tilde\theta_\rho(-x)]} 
   \right],
\end{eqnarray}
where $\tilde v_\rho=v_\rho/\cosh2\phi_\rho$ and
$K_\rho=e^{2\phi_\rho}$.  Similarly, for the spin Hamiltonian, we get
\begin{eqnarray}
   \hat H_\sigma &=& \int_{-l}^l dx \left[\pi\tilde v_\sigma\,
   (\partial_x\tilde\theta_\sigma(x)/2\pi)^2 \right.
\nonumber \\
\label{spin}
   &+& \left. \frac{2g_{1\perp}}{(2\pi\zeta)^2}\,
   e^{-i\sqrt{2K_\sigma}\,[\tilde\theta_\sigma(x)
   -\tilde\theta_\sigma(-x)]} \right],
\end{eqnarray}
where $\tilde v_\sigma=v_\sigma/\cosh 2\phi_\sigma$ and
$K_\sigma=e^{2\phi_\sigma}$.

The umklapp term [the last term in Eq.\ (\ref{charge})] is irrelevant
when $K_\rho>1$ \cite{Firsov}.  In the opposite case $K_\rho<1$, this
term is relevant and leads to opening of an energy gap in the charge
sector.  At the particular value $K_\rho=1/2$, the charge Hamiltonian
$\hat H_\rho$ can be mapped to a model of noninteracting fermions with
an energy gap \cite{Firsov,Luther}.  By introducing the effective
fermions (holons)
\begin{equation}
  \tilde\psi_\rho(x)=e^{i\tilde\theta_\rho(x)}\,(2\pi\zeta)^{-1/2},
\end{equation}
and taking into account the commutation relations (\ref{commutator}),
the last term in Eq.\ (\ref{charge}) can be written as
\begin{equation}
   \frac
   {e^{-i[\tilde\theta_\rho(x)-\tilde\theta_\rho(-x)]}}
   {(2\pi\zeta)^2}
   =-i\,{\rm sign}(x)
   \frac{\tilde\psi_\rho^\dagger(x)\tilde\psi_\rho(-x)}{2\pi\zeta}.
\label{sign}
\end{equation}
Then the charge Hamiltonian (\ref{charge}) becomes
\begin{eqnarray}
  \hat H_{\rho} &=& \int_{-l}^l dx \left\{ -\tilde v_\rho\,
    \tilde\psi_\rho^\dagger(x)\,i\partial_x\tilde\psi_\rho(x) \right.
\nonumber \\
  &-& \left. [g_3^b-ig_3^a\,{\rm sgn}(x)] \,
    \tilde\psi_\rho^\dagger(x)\tilde\psi_\rho(-x) /2\pi\zeta \right\}.
\label{H_rho}
\end{eqnarray}

The energy spectrum $\varepsilon_\rho$ of the Hamiltonian
(\ref{H_rho}) can found from the corresponding Schr\"odinger equation
\FL
\begin{equation}
\label{shreq}
   -i\tilde v_\rho\partial_x\tilde\psi_\rho(x)-
   \frac{g_3^b-ig_3^a{\rm sgn}(x)}{2\pi\zeta}
   \tilde\psi_\rho(-x)=\varepsilon_\rho\tilde\psi_\rho(x).
\end{equation}
Eq.\ (\ref{shreq}) is formally equivalent to Eq.\ (\ref{onlyright})
with $g_3^b$ and $g_3^a$ playing the roles of $\bar\mu$ and $\bar t$,
respectively.  However, as Eq.\ (\ref{g_3}) shows, $g_3^b\propto\bar
t$ and $g_3^a\propto\bar\mu$, so the roles of $\bar\mu$ and $\bar t$
are reversed in Eq.\ (\ref{shreq}) compared with Eq.\ 
(\ref{onlyright}).  This is a consequence of the factor $-i\,{\rm
  sign}(x)$ in Eq.\ (\ref{sign}).  Eq.\ (\ref{shreq}) gives the
spectrum $\varepsilon_\rho=\sqrt{\tilde v_\rho^2 k^2+\Delta_\rho^2}$
for the bulk states with the charge gap
$\Delta_\rho=\sqrt{(g_3^b)^2+(g_3^a)^2}/2\pi\zeta$.  It also gives an
edge solution localized around $x=0$:
\begin{eqnarray}
   && \tilde\psi_{\rho e}(x) = e^{-\kappa\,|x|}, 
\label{psi-e} \\
   && \varepsilon_{\rho e} = -g_3^b/2\pi\zeta
   =-\Delta_\rho\,\bar t/\sqrt{\bar t^2+2\bar\mu^2}, 
\label{E_rho-e} \\
   && \kappa = -g_3^a/2\pi\zeta\tilde v_\rho
   = (\Delta_\rho/\tilde v_\rho)\,
   \bar\mu/\sqrt{\bar\mu^2+\bar t^2/2},
\label{kappa-e}
\end{eqnarray}
where we used Eq.\ (\ref{g_3}) assuming $U>0$ for the repulsive
Hubbard model.  We see that, opposite to the electron edge state
(\ref{E_e}), the energy (\ref{E_rho-e}) of the holon edge state
(\ref{psi-e}) is determined by the bond dimerization $\bar t$ and its
localization length (\ref{kappa-e}) is controlled by the site
dimerization $\bar\mu$.  The holon edge state exists only when
$g_3^a<0$, i.e.\ $\bar\mu>0$.  Taking into account Eq.\ (\ref{t-mu}),
we conclude that the holon edge state exists when the chemical
potential of the edge site ($n=1$) is higher, i.e.\ when the anion
displaces toward the edge site, as shown in Fig.\
\ref{fig:chain}. Thus, we predict that the holon edges states should
appear in (TMTTF)$_2$X below the anion transition temperature
\cite{Monceau}.  Their energies (\ref{E_rho-e}) depend on whether the
first bond at the edge is short or long.

Similar consideration \cite{Fabrizio} can be done for the spin
Hamiltonian (\ref{spin}).  A gap in the spin sector opens when
$K_\sigma<1$, i.e.\ $g_{1\|}<0$ \cite{Firsov}, whereas the required
condition for the existence of an edge state is $g_{1\perp}>0$.
However, $g_{1\perp}=g_{1||}$ in a physical system because of the spin
SU(2) invariance.  So, it is not possible to satisfy the conditions
for opening of a gap and existence of an edge state simultaneously.
Thus, we conclude, contrary to Ref.\ \cite{Fabrizio}, that spinon edge
states cannot exist.

In conclusion, we have shown that holon edge states may appear in a 1D
model of interacting electrons with an energy gap in the charge
sector.  The existence of the holon edge state and its localization
length are controlled by the real part of the umklapp interaction
amplitude, whereas its energy is determined by the imaginary part.  In
the quasi-one-dimensional organic conductors (TMTTF)$_2$X, these
umklapp amplitudes are proportional to the site and bond
dimerizations, correspondingly.  The bond dimerization is always
present in these materials.  However, the site dimerization occurs
only below the recently discovered ferroelectric transition, where the
inorganic anions X displace to asymmetric positions \cite{Monceau}.
We predict that the holon edge states should appear below the anion
transition.  They could be probed by electron tunneling into the ends
of 1D chains or by other types of spectroscopy.
  
V.M.Y.\ is grateful to K.~Le~Hur and S.~A.~Brazovskii for drawing his
attention to Refs.\ \cite{Fabrizio,Gogolin} and \cite{Monceau}
respectively.  This work was supported by the Packard Foundation and
NSF Grant No.\ DMR-9815094.

\vspace{-1.5\baselineskip}


\begin{references}

\vspace{-4.5\baselineskip}

\bibitem{Davison} S. G. Davison and M. St\c{e}\'slicka, {\it Basic
    Theory of Surface States} (Oxford University Press, Oxford, 1996).

\bibitem{Hu} C.-R. Hu, Phys. Rev. Lett. {\bf 72}, 1526 (1994).
  
\bibitem{Yakovenko-SrRu} K. Sengupta, H.-J. Kwon, and V. M. Yakovenko,
  cond-mat/0106198.
  
\bibitem{TMTSF} TMTSF stands for tetramethyltetraselenafulvalene, X
  for inorganic anions such as $\rm PF_6$ and $\rm AsF_6$, and TMTTF
  for tetramethyltetrathiafulvalene.
  
\bibitem{Sengupta-SC} K. Sengupta, I. \u{Z}uti\'c, H.-J. Kwon, V. M.
  Yakovenko, and S. Das Sarma, Phys. Rev. B {\bf 63}, 144531 (2001).

\bibitem{Sengupta-FISDW} K. Sengupta, H.-J. Kwon and V. M. Yakovenko,
  Phys. Rev. Lett. {\bf 86}, 1094 (2001).

\bibitem{Firsov} Yu. A. Firsov, V. N. Prigorin, and Chr. Seidel,
  Physics Reports {\bf 126}, 245 (1985).

\bibitem{Fabrizio} M. Fabrizio and A. O. Gogolin, Phys. Rev. B {\bf
    51}, 17827 (1995).
  
\bibitem{Gogolin} A. O. Gogolin, Phys. Rev. B {\bf 54}, 16063 (1996);
  K. Le Hur, Europhys. Lett. {\bf 49}, 768 (2000).

\bibitem{Yamaji} T. Ishiguro, K. Yamaji, and G. Saito, {\it Organic 
    Superconductors} (Springer, Berlin, 1998). 
  
\bibitem{Monceau} F. Nad, P. Monceau, C. Carcel, and J. M. Fabre,
  J. Phys.  Cond. Matt. {\bf 12}, L435 (2000); D. S. Chow {\it et
  al.}, Phys.  Rev. Lett. {\bf 85}, 1698 (2000); P. Monceau,
  F. Ya. Nad, and S. Brazovskii, Phys. Rev. Lett. {\bf 86}, 4080
  (2001).

\bibitem{Braz-polymer} S. A. Brazovskii, Zh. Eskp. Teor. Fiz. {\bf
  78}, 677 (1980) [Sov. Phys. JETP {\bf 51}, 342 (1980)];
  S. A. Brazovskii, N. N. Kirova, and S. I. Matveenko,
  Zh. Eksp. Teor. Fiz. {\bf 86}, 743 (1984) [Sov. Phys. JETP {\bf 59},
  434 (1984)].
  
\bibitem{Luther} A. Luther and V. J. Emery, Phys. Rev. Lett. {\bf 33},
  589 (1974).

\end{references}
\end{document}